\documentclass[aps,prd,twocolumn,groupedaddress,showpacs,preprintnumbers,draft]{revtex4}
\usepackage{epsf} 
\begin{document}
\preprint{BROWN-HET-1380}
\preprint{MCGILL-22-03}
\preprint{SU-GP-03/10-2}
\preprint{\tt hep-th/0310368}
\title{Spectral Dependence of CMB Polarization and Parity}
\author{K.R.S. Balaji}%
	\email[Email:]{balaji@hep.physics.mcgill.ca}
\affiliation{Department of Physics, McGill University, Montr\'eal, Qu\'ebec, 
H3A 2T8, Canada}
\author{Robert H. Brandenberger}%
	\email[Email:]{rhb@het.brown.edu}
\affiliation{Department of Physics, Brown University, Providence, RI 02912, 
USA} 
\author{Damien A. Easson}%
	\email[Email:]{easson@physics.syr.edu}
\affiliation{Department of Physics, Syracuse University, Syracuse, NY 13244-1130, USA}

\date{\today}

\begin{abstract}
The polarization of the cosmic microwave background radiation (CMBR) can 
serve as a probe for nonstandard parity violating interactions. Many such 
interactions are predicted in particle physics models arising from theories
with extra dimensions such as superstring theory. These interactions produce 
an optical activity that depends on the space-time nature of the 
parity violating field. In particular, it is possible to obtain a 
frequency-dependent differential rotation of the polarization axis. The form of the
frequency dependence is sensitive to the time evolution of the 
parity odd background field. Thus, one may be able to study
a broad class of parity violating operators 
through polarization measurements of the CMBR.
\end{abstract}

\pacs{98.80.Cq.}

\maketitle

\def\Box{\nabla^2}  
\def\ie{{\em i.e.\/}}  
\def\eg{{\em e.g.\/}}  
\def\etc{{\em etc.\/}}  
\def\etal{{\em et al.\/}}  
\def\S{{\mathcal S}}  
\def\I{{\mathcal I}}  
\def\mL{{\mathcal L}}  
\def\H{{\mathcal H}}  
\def\M{{\mathcal M}}  
\def\N{{\mathcal N}} 
\def\O{{\mathcal O}} 
\def\cP{{\mathcal P}} 
\def\R{{\mathcal R}}  
\def\K{{\mathcal K}}  
\def\W{{\mathcal W}} 
\def\mM{{\mathcal M}} 
\def\mJ{{\mathcal J}} 
\def\mP{{\mathbf P}} 
\def\mT{{\mathbf T}} 
\def\mR{{\mathbf R}}
\def\mS{{\mathbf S}}
\def\mX{{\mathbf X}}
\def\mZ{{\mathbf Z}}
\def\eff{{\mathrm{eff}}}  
\def\Newton{{\mathrm{Newton}}}  
\def\bulk{{\mathrm{bulk}}}  
\def\brane{{\mathrm{brane}}}  
\def\matter{{\mathrm{matter}}}  
\def\tr{{\mathrm{tr}}}  
\def\normal{{\mathrm{normal}}}  
\def\implies{\Rightarrow}  
\def\half{{1\over2}}  
\newcommand{\da}{\dot{a}}
\newcommand{\db}{\dot{b}}
\newcommand{\dn}{\dot{n}}
\newcommand{\dda}{\ddot{a}}
\newcommand{\ddb}{\ddot{b}}
\newcommand{\ddn}{\ddot{n}}
\def\be{\begin{equation}}
\def\ee{\end{equation}}
\def\bea{\begin{eqnarray}}
\def\eea{\end{eqnarray}}
\def\bs{\begin{subequations}}
\def\es{\end{subequations}}
\def\g{\gamma}
\def\G{\Gamma}
\def\vp{\varphi}
\def\mpl{M_{\rm P}}
\def\ms{M_{\rm s}}
\def\ls{\ell_{\rm s}}
\def\lp{\ell_{\rm pl}}
\def\l{\lambda}
\def\gs{g_{\rm s}}
\def\d{\partial}
\def\co{{\cal O}}
\def\sp{\;\;\;,\;\;\;}
\def\spa{\;\;\;}
\def\r{\rho}
\def\dr{\dot r}
\def\dt{\dot\varphi}
\def\e{\epsilon}
\def\k{\kappa}
\def\m{\mu}
\def\n{\nu}
\def\om{\omega}
\def\tn{\tilde \nu}
\def\p{\phi}
\def\vp{\varphi}
\def\P{\Phi}
\def\r{\rho}
\def\s{\sigma}
\def\t{\tau}
\def\x{\chi}
\def\z{\zeta}
\def\a{\alpha}
\def\b{\beta}
\def\de{\delta}
\def\bra#1{\left\langle #1\right|}
\def\ket#1{\left| #1\right\rangle}
\newcommand{\stt}{\small\tt}
\renewcommand{\theequation}{\arabic{section}.\arabic{equation}}
\newcommand{\eq}[1]{equation~(\ref{#1})}
\newcommand{\eqs}[2]{equations~(\ref{#1}) and~(\ref{#2})}
\newcommand{\eqto}[2]{equations~(\ref{#1}) to~(\ref{#2})}
\newcommand{\fig}[1]{Fig.~(\ref{#1})}
\newcommand{\figs}[2]{Figs.~(\ref{#1}) and~(\ref{#2})}
\newcommand{\GeV}{\mbox{GeV}}
\def\ricci{R_{\m\n} R^{\m\n}}
\def\riemann{R_{\m\n\l\s} R^{\m\n\l\s}}
\def\triemann{\tilde R_{\m\n\l\s} \tilde R^{\m\n\l\s}}
\def\tricci{\tilde R_{\m\n} \tilde R^{\m\n}}
\section{Introduction}  
Perhaps the single most important source of observational data in cosmology 
is the cosmic microwave background radiation (CMBR). The observed radiation 
was last scattered around a red-shift of $z=1100$ (the time of ``recombination'') 
and carries a tremendous amount of information about the physics of the 
early universe. It is expected that through such cosmological data 
extensions of new physics (normally insensitive to the electroweak scale)
such as string theory or any other alternative theories may establish 
contact with observation. 

The spectral distribution of the CMBR is to very high accuracy \cite{COBE}
that of a black body. Primordial density fluctuations seed anisotropies
in the CMBR, and the physics at the surface of recombination generates a
polarization of these anisotropies (for a review see e.g. \cite{Huprimer}).
In this Letter, we point out that many parity-odd terms in the effective
Lagrangian for low energy physics lead to a frequency-dependent
rotation of the polarization axis. In fact, the details of the frequency
dependence of this rotation may give insight into the nature of the
parity-odd terms, and therefore provide an avenue to probe the high
energy fundamental physics which is responsible for the parity-odd
terms in the low energy effective action.

Our analysis builds on early work by Carroll, Field and Jackiw 
who pointed out that parity-odd terms in the effective Lagrangian,
e.g. a Chern-Simons term \cite{CFJ} or a term coupling a new scalar
field $\phi$ (such as the scalar fields used to model
quintessence) to $F \wedge {}^{*}F$ \cite{CF} leads to the
rotation of the polarization axis of the radiation from distant quasars. 
Carroll then studied the bounds on quintessence models coming from the
observational bounds on this effect \cite{Carroll}. Lue, Wang and Kamionkowski
\cite{LWK} (and independently Lepora \cite{Lepora:1998ix}) pointed out that the coupling 
terms mentioned above will
also produce a rotation in the polarization axis of the CMBR (``cosmological
birefringence'') and that this effect
can be used to probe parity-violating interactions. The new
aspects of our work are the focus on the frequency dependence of the effect,
and the application of this result to the study of fundamental physics. The frequency dependence of 
the birefringence discussed here is different from that produced by Faraday rotation,
and hence can be tested experimentally.
 
In order for new terms ${\cal L}_i$ in the effective Lagrangian to produce
cosmological birefringence, the terms in the fundamental
theory leading to ${\cal L}_i$ must possess electromagnetic (EM) interactions. 
To be specific, let us consider parity-odd Chern-Simons (CS) terms of a given 
form. A couple of examples are (i) a three-form Kalb-Ramond field 
(see e.g. \cite{sjgates}) which is ubiquitous in string models and 
(ii) a one-form pseudo-scalar field \cite{CFJ}; both of these can have 
electromagnetic interactions. 

It is interesting to study the nature of the parity
violating interactions by looking at the qualitative and quantitative
features of the cosmological birefringence. For instance, what multiplies
${}^{*}F$ in the new term ${\cal L}_i$ of the effective Lagrangian
can be a form $F_j$ of various dimensionalities, not just a one form as assumed
in previous work. Can one determine the nature of this form $F_j$ from the
specifics of the frequency dependence? In this Letter, we
neglect the space-dependence of $F_j$ (which is well justified to leading
order in our context of cosmology). We show that the 
predicted birefringence has a spectral dependence with signatures 
that depend on the form $F_j$. In particular, the time dependence of the
underlying fields plays a crucial role. 

We conclude that the polarization of the CMBR can be 
sensitive to any derivative couplings in the parity-odd operator in the
effective Lagrangain. Since a frequency dependence of the
rotation of the polarization axis is predicted in many cases and since it
carries information about the underlying fundamental physics,
we propose that it is of interest to perform a spectral analysis of any 
birefringently polarized CMBR. 

\section{A general analysis}
Our starting point consists of adding to the
Maxwell equations a term coming from the interaction of the
EM gauge field $A_{\mu}$ with an arbitrary parity odd CS field
$\O_\m$. The interaction Lagrangian is given by
\be
\mL = \O_\m A_\n \tilde F^{\m\n}~,
\label{cs}
\ee
where $\tilde F^{\m\n}$ is the dual of the usual antisymmetric EM field tensor.
The resulting field equations are \cite{CFJ}
\be
\nabla_\m F^{\m\n}= (\O \cdot \tilde F)_\n~;~\nabla_\m \tilde F^{\m\n}=0
\,.
\label{gen1}
\ee
Due to gauge invariance, $\O$ must be curl free and hence
expressible in terms of a gradient of a scalar. In general, 
it is either of the type
\be
\O_\mu \sim H_{\m\n} \d_\n\phi \, \label{single}~,
\ee 
or a term with more derivatives 
\bea
\O_\mu &\sim& H_{\mu\l\sigma...n} [\d_\l \phi \wedge \d_\sigma \phi...
\wedge \d_n\phi] \,\,\, {\rm or} \nonumber\\ 
\O_\mu &\sim& H_{\mu\l\sigma...n} [\d_\l \phi \d_\sigma \phi...
\d_n\phi] \, . \label{higher}
\eea
Here, $H$ denotes a tensor field (in general independent of the fields
which we have already introduced) which is either symmetric or
antisymmetric and can have space-time dependence, 
$H \equiv H(\eta, x)$. In the simplest case of (\ref{single}), 
$\O_{\mu}$ is the covariant derivative of a 
scalar field like the axion \cite{sikivie}. 

The physical process we focus on in this Letter which arises
due to the above new parity-odd
interaction term is cosmological birefringence, 
the rotation of the plane of polarization of the CMBR. Another effect
(to be discussed elsewhere) is a temperature and polarization anisotropy 
due to a fluctuating $H$ field.

Let us analyze the consequence of (\ref{gen1}) when 
$\O$ is a spatially homogeneous
operator with $H \equiv H(\eta)$. 
We choose a spatially flat Friedmann-Robertson-Walker background metric 
$ds^2 = a^2(\eta)(-d\eta^2 + dx_idx^i)$
where $a(\eta)$ is the scale factor of the Universe and $\eta$ is 
conformal time. In this case (\ref{gen1}) translates to
\bea
\frac{\d}{\d\eta}[a^2(\eta)\vec E]-\nabla \times
[a^2(\eta)\vec B] &=& -2C(\frac{d\Phi}{d\eta})^n [a^2(\eta)\vec B]~,\nonumber\\
\frac{d\Phi}{d\eta}=H^{1/n} \frac{d\phi}{d\eta}&\equiv& \Phi'~.
\label{gen2}
\eea
Here, $n$ denotes the number of derivatives in $\O$, and the constant $C$ 
(used to specify the dependence in (\ref{single}) and (\ref{higher})) is 
determined by the mass scale $M$ characteristic of the physics leading to
the effective Lagrangian (\ref{cs}), i.e. $C \sim M^{1-d}$ 
where $d$ is the dimension of the fields of the operator $\O$ appearing in
(\ref{higher}). 
Setting $F_\pm = a^2(\eta)\vec B_\pm (\eta)$, leads to the differential
wave equation
\be
\frac{d^2 F_\pm}{d \eta^2}+(k^2 \mp 2k C\Phi'^n)F_\pm =0\,~.
\label{weq1}
\ee

If $\Phi' \neq 0$, the left- and right-circular modes ($\mp$ states) 
of the photon are split, leading to a modified dispersion relation  
\be
\omega_\pm^2 = k[k \mp 2C(\Phi')^n]~.
\label{weq2}
\ee
Therefore, 
if the photons travel
for a time interval $\delta\eta$, they obtain a differential
rotation (birefringence) given by 
\be
\delta \theta = C(\phi^{\prime})^n H(\eta)\delta\eta + 
C\delta H(\eta) (\phi^{\prime})^n\delta\eta~.
\label{phn}
\ee 
The first term in (\ref{phn}) arises due to the presence of the $\phi$ field
while the second term arises due to the variation in $H(\eta)$ over the time
slice $\delta\eta$. The latter term also gives rise to a change in the
photon energy as can be seen from (\ref{weq2}). Therefore, the
differential rotation $(d(\delta\theta)\equiv d\theta)$ is correlated to the 
change in the photon energy. This effect is purely due to the time dependence of 
$H(\eta)$. Thus, the resulting optical birefringence can become a function of the 
frequency change, an effect which to our knowledge
has not been pointed out (see e.g. \cite{LWK}).
We note that this effect can also occur if the $\phi$ field is a 
temporally oscillating massive field while $H(\eta)$ is constant. This 
possibility can be trivially examined along the lines of the present analysis.
As described in (\ref{phn}), it is important to observe that the 
rotation leading to a change of the photon 
energy is an additional contribution to an existing rotation due to the 
$\phi$ field.
Note also, that according to (\ref{phn}), the  
rotation angle $\theta$ depends on the amplitudes of both 
the $\phi$ and $H$ fields. Since it depends linearly on the constant
$C$, it follows that the
birefringence due to higher dimensional operators is suppressed 
by the corresponding powers of $M$.
Assuming that the field $\phi$ is massless, we have (neglecting the expansion
of the Universe)
\be \label{mlphi}
\phi(\eta) = {\dot \phi}_0 \cdot \eta + \phi_0 \, ,
\ee
where ${\dot \phi}_0$ and $\phi_0$ are constants. For this case, 
substituting for $C$, the rotation angle (\ref{phn}) becomes
\be
\theta (\eta) = [H(\eta) (\phi^{\prime})^n M^{1-d}]\eta~.
\label{rphn1}
\ee

\section{Spectral and Time Dependence}

In this section, we study the frequency dependence of the
cosmological birefrigence identified in the previous section.

\subsection{Effects for $\eta_ < \eta_{rec}$}

First, let us elaborate on the electromagnetic effects that occur for all 
interaction times $\eta$ before the time of recombination $\eta_{rec}$. 
These are not actually observable today. In the next subsection we 
discuss the conversion to effects observable 
at  times $\eta \geq \eta_{rec}$. 

Let us begin by clarifying what we mean by spectral dependence of 
the differential 
birefringence in (\ref{phn}). Physically, due to the presence of the 
time-dependent background field, 
the photon undergoes two simultaneous effects:
\begin{itemize}
\item{(i)} an optical birefringence due to 
the $H(\eta)$ field (along with the $\phi$ field), and 
\item{(ii)} a variation of the photon
energy due to its interaction with the time varying $H(\eta)$ field.
\end{itemize}
The second feature is seen by Taylor expanding (\ref{weq2}) 
\be \label{eloss}
\frac{d \omega_\pm}{d\eta} \approx \mp C (\frac{d\phi}{d\eta})^n\cdot
\frac{d H(\eta)}{d\eta}\neq 0~.
\label{approx}
\ee
In the above, we have assumed that the field $\phi$ is massless, and
thus (\ref{mlphi}) holds. If this were not the case, and the
time dependence of $\phi$ were nonlinear, then there would be an
extra term in (\ref{eloss}) proportional to $H$ and to the second
derivative of $\phi$.
We stress that the above change in energy is due to interactions at all times 
$\eta< \eta_{rec}$.
Clearly, if the background field were constant, then a change in the 
rotation would not cost the photon any energy. In other words, the
scattering leading to the optical birefringence would be an elastic process. 
Then, the rotations described here would be frequency independent.  
For an inealstic process, the change in the photon energy $(d\omega_\pm)$ 
associated with a differential rotation $(d\theta)$ is
\be
\frac{d\omega_\pm}{d \theta} = \frac{d\omega_\pm}{d \eta} \cdot \frac{d \eta}
{d\theta}~\mbox{for ~all}~\eta \leq \eta_{rec} \, ,
\label{energy1}
\ee
where all derivatives are taken at constant value of the wavenumber $k$.
Using (\ref{phn}) and (\ref{energy1}) we find
\be
\frac{d\omega_\pm}{d \theta}= \mp \frac{H^\prime}{H^\prime \eta + H}~;~
H^\prime = \frac{d H(\eta)}{d \eta}~.
\label{energy2} 
\ee

The above result allows us to infer the time dependence of $H(\eta)$ 
by inverting (\ref{energy2}). We obtain
\be
\frac{d H(\eta)}{H(\eta)} = \frac{f_\pm d\eta}{1 \mp f_\pm \eta}~;~f_\pm \equiv
\frac{d\omega_\pm}{d\theta}~.
\label{heta1}
\ee
Integrating (\ref{heta1}) leads to
\be
\log \Big(\frac {H(\eta)}{M_X}\Big ) =  \int \frac{f_\pm d\eta}
{1 \pm \eta f_\pm }~:~\eta < \eta_{rec}.
\label{heta2}
\ee
Here, $M_X$ can be identified with the 
scale at which the electromagnetic interaction is defined.

\subsection{Observable effects for $\eta_\geq \eta_{rec}$ }

What is actually observable today is a spectral/frequency dependence in
the rotation angle. In principle, this dependence is a correlation of an 
observable energy width to a given change in the rotation angle. 
The actual observable change in rotation is
\be
d\tilde \theta = 
\frac{d \theta}{d\omega_\pm} \cdot d\omega_\pm = f_\pm^{-1} d\omega_\pm  
~\mbox{for ~all}~\eta >\eta_{rec} \, .
\label{obs1}
\ee
Note that the effect vanishes if the scattering process is elastic, i.e. if 
$d\omega_\pm = 0$.
The tilde in (\ref{obs1}) denotes the actual observable rotation. 
Therefore, for a measurable frequency width 
$\delta\tilde\omega$, the true spectral dependent differential rotation is
\be
\frac{d \tilde \theta} {d\tilde \omega}= f_\pm ^{-1} 
\cdot \frac{d \omega_\pm}{d\tilde\omega}~.
\label{obs2}
\ee
As a consistency check, (\ref{obs2}) indicates that if no 
spectral dependence of the differential rotation $(d\omega_\pm \to 0)$ 
were observed, we would deduce from (\ref{heta2}) that
$H(\eta)$ is a constant field $\sim M_X$. In order 
to obtain a frequency dependent birefringence one requires a  background field
which is time dependent.
Finally, using (\ref{weq2}) and (\ref{obs2})
\be
\frac{d \tilde \theta} {d\tilde \omega}= \frac{C(\phi^{'})^n}{d\tilde\omega}
\cdot \int_{\eta}^{\eta_{rec}} f_\pm ^{-1} \cdot \frac{d H(\eta)}{d\eta}d\eta
\, ,
\label{obs3}
\ee
where the integration runs over a finite short time interval from $\eta$
to $\eta_{rec}$ during which the photons are scattering.

We illustrate the above results for a specific choice of $H(\eta)$,
namely $H(\eta) \sim h\cdot \eta$. This could arise (neglecting the
expansion of the Universe) if $H$ were a massless field obeying the 
equation $\Box H(\eta) =0$, and taking $H$ to be spatially homogeneous.
In the linearized approximation, using (\ref{energy2}), we find that,
for an infinitesimal
energy interval $\Delta\omega$, the left and right modes rotate by an amount
\be
\Delta \theta_\pm = \mp 2\eta\Delta \omega_\pm ~\Rightarrow f_\pm^{-1} = \mp 
2\eta~.
\label{linear1}
\ee
This indicates that the differential rotation 
(for $\eta < \eta_{rec}$) grows with 
energy and is independent of the density of the background field. For this
scenario, we find from (\ref{obs3}) the observable differential rotation to be
\be
\frac{d \tilde \theta} {d\tilde \omega}= \mp\frac{Ch(\phi^{\prime})^n}{2d\tilde\omega}
\eta_{rec}^2~.
\label{ex1}
\ee

Note that the observable spectral dependence vanishes for 
$\eta = \eta_{rec}$ since in this case there are no scattering events
that lead to a birefringence. Naively, it also appears from (\ref{ex1})
that in order to obtain a large and thus observable effect, the time width 
in scattering $(\eta - \eta_{rec})$ needs to be large. However, in
this case our neglect of the expansion of the Universe is no longer
justified (this could easily be corrected by taking into account the
Hubble friction in the equation of motion for $H$ and using the
resulting time dependence). More importantly, the incoherence of
the scattering events would need to be taken into account, and this would
alter the amplitude of the predicted signal. However, on the positive
side, we could easily retain a signal from physics in the very early
Universe, at times $(\eta \ll \eta_{rec})$, have the field $H(\eta)$
depend on time in the very early Universe but be frozen in the late
Universe. In this case, there would be no cosmological constraints on
the scenario coming from demanding that the kinetic energy of the
$H$ field be negligible in the late Universe. 

Let us conclude this section by evaluating the order of the magnitude
of the differential birefringence in the case $n = 1$. Taking 
$\phi$ to be dimensionless, and $H$ to have the canonical dimension 1,
we obtain
\be \label{estimate}
\frac{d \tilde \theta} {d\tilde \omega} \sim 
{{M_2} \over M} {{\eta_{rec}^2} \over {\eta_1 \eta_2}} {1 \over {d\tilde \omega}} \, ,
\ee
where $M_2$ is the amplitude of $H$ and $\eta_1$ and $\eta_2$ are the time
scales of the change in $\phi$ and $H$, respectively. Here, we have set
the dimensionless amplitude of $\phi$ to 1. For massless fields,
we expect the time scales are given by the Hubble time, and thus the
ratio of times in (\ref{estimate}) is of the order one. Hence, logarithmic 
differential birefringence can be of the order 1
if $M_2 \sim M$ when $\phi$ is measured in units where $\phi = 1$ at
$\eta = \eta_{rec}$. In turn, this will be the case if the new physics
only contains one new mass scale. 

One may be concerned that the polarization signature
could be washed out by Compton scattering while the photons are in 
thermal equilibrium 
\footnote{We thank Guy D. Moore for raising this concern.}. 
However, note the potential depolarization
effects due to Compton scattering only compete with the cosmic 
birefringence at early times when the photons are in thermal
equilibrium. The depolarization effects rapidly decrease at later times.  
Now, there is a large time interval between when the photons fall out of
equilibrium and when they last scatter, and it is during
this time interval (which corresponds to the finite thickness
of the last scattering surface) that our polarization
effect can build up. 

Demanding that the kinetic energy in $\phi$ and $H$ not destroy
the successful predictions of nucleosynthesis will lead to 
bounds on the ratio of masses (typically, $M_2$ will have to be
small compared to $M$ in order that the condition is satisfied).
However, the details of the bounds will depend on the model in
several ways, e.g. on the form of the kinetic energy term for
the new fields, the history of the time dependence of the field, etc.. 
We do not see any obvious reason why our effect should be
so small as to be unmeasurable in upcoming polarization experiments.

\section{A string based model}

The goal of this section is to demonstrate that interaction terms of the
type postulated at the beginning of this paper arise
in theories of fundamental physics. 
To be specific, consider the relevant parts of a
ten dimensional type-IIB superstring bosonic action (see
e.g. \cite{cliff} for a recent review). In addition to the
terms in the action coming from the graviton, the dilaton
and the NS-NS \footnote{Here, NS stands for Neveu-Schwarz, and
R-R will later be used to denote Ramond-Ramond.}
two form (these terms give the action for
dilaton gravity with the addition of a term involving the
square of the field strength of the NS-NS two form), there
are terms of the form
\begin{eqnarray}
S_{\rm IIB} &=& -\frac{1}{2\k_0^2} \int\!d^{10}\!x \sqrt{-g}
\left\{\frac{1}{12} (G^{(3)} + C^{(0)} B^{(3)})^2 \right\}\nonumber\\
&+&{1\over8\k_0^2}\int \!d^{10}\!x B^{(2)}C^{(2)}G^{(3)}\,B^{(3)} \, ,
\label{sugra}
\end{eqnarray}
where we have adopted the usual notation. Thus,
$G^{(3)}=dC^{(2)}$ is the field strength of a R-R two form
$C^{(2)}$, $C^{(0)}$ is a zero-form R-R scalar field while $B^{(3)}$ is the 
field strength of the NS-NS two form. 
In what follows we will assume that the dilaton is fixed. 

Consider the first term in (\ref{sugra}). The four-dimensional 
reduced action giving the kinetic term for the $B$ field is
\be\label{sh}
S = \frac{1}{12} \int d^4 \! x \, \sqrt{-g} \, (C^{(0)})^2\, 
B_{\m\n\a}B^{\m\n\a} \, ,
\ee
where $B_{\m\n\a} = \d_{[\m}B_{\n\a]}$, and 
$B_{\m\n}$ is the Kalb-Ramond field. 
In the context of cosmology and, in particular,
in the case of the Pre-Big-Bang scenario (for a recent review see 
\cite{Gasperini:2002bn}) the role of such a term has been studied, see
e.g. \cite{Kaloper:1991rw}. Fixing the scalar field $C^{(0)}$, the 
equation of motion becomes
\be
\d_\n ( \sqrt{-g} \, B^{\m\n\a} )= 0
\ee
and is satisfied by introducing the dual axion field $\sigma$,
\be
B^{\m\n\a}=\frac{1}{\sqrt{-g}C} \epsilon^{\m\n\a\b}\d_\b \sigma~.
\ee
Then, the action (\ref{sh}) becomes
\be
S =  \frac{1}{2} \int d^4 \! x \, \sqrt{-g} \, g^{\a\b} 
\d_\a \s \d_\b \s \,.
\ee

The electromagnetic interaction is obtained through the 
interference term $(G^{(3)} B^{(3)})$, 
which gives the low energy four dimensional action
\be
S =  \frac{1}{2} \int d^4 \! x \, \sqrt{-g} \, g^{\a\b} 
 B^{\m\n\a}A_\m \tilde F_{\n\a}~.
\label{s1}
\ee
It can be easily shown that for (\ref{s1}) the operator 
$\O_\mu$ corresponds to a
space-time constant $H$ field with $\O_\m = \d_\m\sigma$. Such an interaction
leads to a frequency independent birefringence.

Next, let us consider the second term in (\ref{sugra}) which
yields the four dimensional effective action
\be
 S =  \frac{3}{4} \int d^4 \! x \, B\d_\n\sigma \Big( A_{2\lambda}
\tilde F^{\n\lambda}_1 - A_{1\lambda}
\tilde F^{\n\lambda}_2 \Big)
 \,.
\label{s2}
\ee
Identifying the linear combination 
$(A_{1\lambda} + A_{2\lambda}=\sqrt{2}A_\lambda)$ 
as the photon field, we obtain the effective action for the 
electromagnetic interaction
\be
 S =  \frac{3}{2} \int d^4 \! x \, B \d_\n\sigma A_\lambda
\tilde F^{\n\lambda} \,.
\label{s3}
\ee

Clearly, (\ref{s3}) reduces to our original action (\ref{cs}) 
with  the identification $\O_\m = H\d_\n\sigma$.
In this case $H\equiv B$ and $\sigma$ are independent 
massless scalar fields, and the $B$ field is taken to be the 
zero-form scalar field which arises after compactifying the 
higher dimensional two-form field $B^{(2)}$.
We note that the model has two massless scalar fields, the axion and 
an additional field $H$. The cosmological significance of the string model 
outlined here essentially follows 
along the lines discussed for the axion field \cite{scosmo}. 

In summary, we have shown that terms in the Lagrangian such as the
ones we have used in our analysis appear quite naturally in
string theory models. This provides a concrete example of how string theory
may ultimately make contact with experiment.

\section{$T-E$ cross correlation and shape}

Spatial field fluctuations lead to density perturbations which in
turn induce CMB temperature anisotropies. Here we show that the
parity-odd interactions discussed in this paper lead to specific
signals in the temperature-polarization cross correlation
function of the CMBR sky \footnote{
The effects are similar to those due to tensor field 
fluctuations \cite{turok}. Indeed, if we assume the $H$ field to be
traceless, the process mimics closely perturbations due to a 
gravitational system
\cite{gravity}.}. 

The cosmic sky can be mapped in terms of temperature $(T)$ and 
polarization $(Q)$ parameters. The polarization can be divided into a
a curl-free piece (the $E$-modes) and a divergence-free piece
(the $B$-modes). The first observation of the 
$E$-mode polarization and the $T-E$ cross correlation was made by the
DASI experiment \cite{dasi}. Recently,
WMAP has produced data for the $T-E$ cross correlation at a much larger
significance \cite{kogut}. These signals result due 
to physics of last scattering (with $z\sim 1000$) and hence 
should be most prominent at smaller angular scales $(\leq 2^o)$. 

Conventionally, the temperature and polarization maps are decomposed
into spherical harmonics. The weight factors of this expansion lead to
nonzero moments $C_l^{XX^\prime}$ (for a multipole $l$). Here, $X$ and 
$X^\prime$ can be $T$, $E$ or $B$, thus yielding auto-correlation 
($X = X^\prime$) and cross-correlation ($X \neq X^\prime$) functions.
In the limit of strict parity invariance, one expects the cross-correlations 
involving $B$ to vanish. However, as in our case, if there is a parity 
violating background
field, then due to the rotation of the photon plane of 
polarization by an amount 
$\theta$ there will result a cross-correlation
\be
C^{TB} = C^{TE}\sin\theta~.
\label{ctb1}
\ee
Therefore, if there is frequency dependent rotation, we find,
\be
\frac{d C^{TB}}{d\tilde\omega_\pm} = - C^{TE} \Big (\frac{d\theta}{d \tilde\omega_\pm}\Big)
\cos\theta ~.
\label{ctb2}
\ee
Thus, we conclude that
the shape of $C^{TB}$ will be the same as that of $C^{TE}$ 
only if the rotations are frequency independent. 
If otherwise, we find an interesting modification which is
determined by the nature of the background field. For instance, as in the case of massless
fields, using (\ref{ex1}) we find this modulation is
\be
\frac{d C^{TB}}{d\tilde\omega_\pm} = \pm C^{TE}\Big (\frac{Ch(\phi^{\prime})^n\eta_{rec}^2}
{2d\tilde\omega}\Big )\cos\theta~.
\label{ctb3}
\ee

\section{Summary}

In summary, we have outlined the possibility of identifying parity odd 
electromagnetic interactions by observations of the CMBR. The
first effect is a cosmological birefringence (rotation of the
polarization axis) which in general will depend on the frequency.
The second effect is a cross-correlation function $C^{TB}$
between the CMB temperature and the B-mode of the polarization,
a correlation function whose frequency dependence is determined
by the cross-correlation function $C^{TE}$ involving
the E-mode of the polarization. The relation between the
two spectra depends on the specifics of the parity odd
interactions. We have seen that a measurement of the differential rotation of 
the photon field can give information about the time dependence of 
the parity odd fields which enter in the interaction terms. In the present 
analysis we demonstrated this for the case of massless time varying fields.

In following the inflationary paradigm of early Universe cosmology, 
we presumed that the parity odd fields $H$ appearing in our
interaction terms are spatially homogeneous. Note that 
in a more general setting where the space-dependence of $H$ is
nontrivial, the spectral dependence of the birefringence is 
sensitive to the spatial distribution of the field. The details
remain to be worked out. 
In the limit of $H$ being constant in space and time, we recover 
from (\ref{phn}) the result that the rotation of the polarization
axis doe not depend on the frequency \cite{LWK,CFJ}. 
The frequency-dependent birefringence discussed here can be distinguished
from the effects of Faraday rotation. The latter effects decrease in
magnitude as the frequency increases 
($\theta_{\rm Faraday}(k) = C \times k^{-2}$, $C$ being a
constant), whereas our cosmological
rotation of the polarization axis increases for larger frequencies.
We anticipate that future CMB  polarization measurements will be tuned 
to measure the spectral dependence of any observable cosmological birefringence 
discussed here.
\begin{acknowledgments}  
We are grateful to Guy D. Moore for discussions and a careful reading of a preliminary
version of this draft. We thank Horace Stoica, Greg Tucker and Mark Trodden for several 
helpful discussions. RB thanks the
McGill theory group for hospitality during visits to Montr\'eal during
which this work was initiated.
RB is supported in part by the US Department of Energy under Contract 
DE-FG02-91ER40688, 
TASK~A. DE is supported in part by NSF-PHY-0094122 and funds provided by 
Syracuse 
University. KB is supported by NSERC (Canada) and by Fonds de recherche sur 
la nature 
et les technologies of Qu\'ebec. 
\end{acknowledgments}
   
\end{document}